\documentclass[preprint2,tighten]{aastex}

\bibpunct{(}{)}{;}{a}{}{,}

\shorttitle{Fossil cluster luminosity function}
\shortauthors{Cypriano, Mendes de Oliveira \& Sodr\'e Jr.}

\def\sol{\mbox{$_{\odot}$}} 
\def\Msol{\hbox{M$_{\odot}$}}

\def\m12{\hbox{$\Delta$m$_{12}$}}
\def\RXJ{RX~J1416.4+2315$\,$}
\def\RXJold{RX~J1552.2+2013$\,$}

\newbox\grsign \setbox\grsign=\hbox{$>$} \newdimen\grdimen
\grdimen=\ht\grsign
\newbox\simlessbox \newbox\simgreatbox
\setbox\simgreatbox=\hbox{\raise.5ex\hbox{$>$}\llap
     {\lower.5ex\hbox{$\sim$}}}\ht1=\grdimen\dp1=0pt
\setbox\simlessbox=\hbox{\raise.5ex\hbox{$<$}\llap
     {\lower.5ex\hbox{$\sim$}}}\ht2=\grdimen\dp2=0pt

\newbox\simppropto
\setbox\simppropto=\hbox{\raise.5ex\hbox{$\sim$}\llap
     {\lower.5ex\hbox{$\propto$}}}\ht2=\grdimen\dp2=0pt

\begin{document}

\title{
Velocity dispersion, mass and the 
luminosity function of the fossil cluster RX~J1416.4+2315
\thanks{Based on observations obtained at the Gemini Observatory,
which is operated by the Association of Universities for Research in
Astronomy, Inc., under a cooperative agreement with the NSF on behalf of
the Gemini partnership: the National Science Foundation (United States),
the Particle Physics and Astronomy Research Council (United Kingdom),
the National Research Council (Canada), CONICYT (Chile), the Australian
Research Council (Australia), CNPq (Brazil) and CONICET (Argentina) --
Observing run ID: GN-2005A-Q-38.}
}

\author{Eduardo S. Cypriano}
\affil{Department of Physics \& Astronomy, University College London,
London, WC1E 6BT}
\email{esc@star.ucl.ac.uk}

\author{Claudia L. Mendes de Oliveira, Laerte Sodr\'e Jr.}
\affil{Departamento de Astronomia, Instituto de Astronomia, Geof\'{\i}sica
e Ci\^encias Atmosf\'ericas da USP, Rua do Mat\~ao 1226, Cidade
Universit\'aria, 05508-090, S\~ao Paulo, Brazil}
\email{oliveira@astro.iag.usp.br, laerte@astro.iag.usp.br}


\begin{abstract}
{
We study the properties of the fossil cluster RX~J1416.4+2315 through g'
and i'-band  imaging and spectroscopy of 25 member galaxies.  The system
is at a mean redshift of 0.137 and has a velocity dispersion of 584 km
s$^{-1}$.  Superimposed onto one quadrant of the cluster field there is
a group of five galaxies at a mean redshift of 0.131, which, if included
as part of the cluster, increases the velocity dispersion to 846 km/s.
The central object of RX~J1416.4+2315 is a normal elliptical galaxy, with
no cD envelope.  The luminosity function of the system, estimated by
the number counts, and statistical background correction, in the range --
22.6 $<$ M$_{g^\prime}$ $<$ --16.6, is well fitted by a Schechter function
with M$_{g^\prime}^* = -21.2 \pm 0.8$ and $\alpha = -1.2 \pm 0.2$ (H$_0$
= 70  km s$^{-1}$ Mpc$^{-1}$, $\Omega_M$=0.3, $\Omega_{\Lambda}$=0.7).
The luminosity function obtained from the spectroscopically confirmed
members in both g$^\prime$ and i$^\prime$  bands agrees 
with the photometric results.  The mass of the system,
$M\sim 1.9 \times 10^{14}$ h$^{-1}_{70}$  M$_\odot$, its M/L of 445 h$_{70}$  
M\sol/L$_B$\sol ~  and
 L$_X$ of 11 $\times$ 10$^{43}$ h$^{-2}_{70}$ ergs s$^{-1}$ (bolometric)
suggest that this system is the second example of known fossil {\it
cluster}, after RX~J1552.2+2013, confirmed in the literature.}

\end{abstract}


\section{Introduction}

  It is now well recognised that a merged group can relax to form a single
elliptical galaxy \citep[e.g.][]{barnes,governato,bode,athanassoula}. Since the
timescale for the luminous group members to coalesce is shorter than the
Hubble time, mergers of groups into elliptical galaxies are indeed expected to
be observed. In 1994, Ponman et al. suggested that the elliptical-dominated
system RX~J1340.6+4018 was probably the remains of what previously constituted
a group, and a similar conclusion was reached by Mulchaey and Zabludoff in 1999
about NGC 1132. It was thought that these galaxies were possibly a merged group
with an X-ray halo with an extent of several hundreds of kpc with a rich
population of dwarf galaxies associated with them. These systems were named
fossil groups. A list of the fossil systems known to date are summarised in
Table 4 of \citet{paper1}.

\citet{vikhlinin99} noted that the space density of fossil groups (which they
called OLEGs, ``X-ray overluminous elliptical galaxies") is comparable to that
of compact groups and that they represent 20\% of clusters and groups with
L$_X$ $>$ 1 $\times$ 10$^{43}\:h_{70}^{-2}$ ergs s$^{-1}$, outnumbering compact
groups of similar X-ray luminosities by a factor 3.5. \citet{jones03} defined a
fossil group as a bound system of galaxies immersed in an extended X-ray halo
brighter than $5 \times 10^{41}\:h_{70}^{-2}$ erg s$^{-1}$, for which the
magnitude difference in the R band between the two brightest galaxies of
thesystem exceeds two magnitudes. Searching for systems which followed these
selection criteria in the WARPS survey, these authors revised the number
density of fossil groups to be $\sim$ 2.4 $\times$ 10$^{-7}$ Mpc$^{- 3}$, or
8--20 per cent of all systems with same X-ray luminosities \citep{jones03}.
Analytical estimates of \citet{chrismiller} found that $\sim$ 5--40\% of groups
could be fossil groups while only { 3--6\%} of more massive systems
($M\sim10^{14} M_\odot$) are expected to be fossil {\it clusters}. The latter
result has been confirmed by a search for fossil clusters within the C4
catalogue of clusters of \citet{chrismiller05}, based on the second release of
the Sloan Digital Sky Survey (SDSS). 

  If the merger interpretation is correct, fossil groups may have seen
little infall of large galaxies since their collapse \citep{donghia04}.
Thus they can be important for studying the formation and evolution
of galaxies and the intragroup medium in isolated systems. They
may also be a link between compact groups and elliptical galaxies
{ \citep{ponman94,jones03}}.
Using N-body/hydrodynamical simulations to investigate the origin of
such systems, \citet{donghia2005} found that fossil groups stand out
from the population of regular groups by their older ages.  Those authors
found that fossil groups
have assembled half of their mass before redshift of one.

 \citet{paper1} recently published
the first secure luminosity function determination of a fossil system,
using spectroscopic data of 36 member galaxies in RX~J1552.2 +2013. Two
unexpected results were (1) the system was not a fossil group
but a fossil cluster, given its richness and its velocity
dispersion (623 km/s) and (2) the luminosity function not only
had a lack of bright galaxies (by selection),
but also had a lack of intermediate-luminosity systems (M$_r'$ = --18 mag).
This was a surprise, since the general case for systems
of similar velocity dispersion is to {\it not} have dips in their 
luminosity functions at intermediate luminosities \citep{propris,popessoII}. 
The exceptions are  the central
regions of rich clusters. For instance, the luminosity function of the
core of the Coma cluster displays a dip 
at R$\sim 17.0$ and B$\sim 18.0$
\citep{Trentham}, which is close to the dip found for the 
luminosity function of the fossil cluster
RX~J1552.2 +2013.

   With the main goal of investigating how unique RX~J1552.2+2013 is,
we have obtained imaging and spectra for the galaxies in
a second system classified by \citet{jones00} as a fossil group:
RX~J1416.4+2315. This object 
has { the highest bolometric X-ray-luminosity (11 $\times$ 
10$^{43}$ h$_{70}^{-2}$ erg\,s$^{-1}$) from all six fossil
groups in the \citet{jones03}
sample. Its extended X-ray emission is centered on the dominant galaxy and
has an elliptical shape whose semi-major axis direction nearly coincides with
the main optical axis of the central galaxy.}
Before this study, this system had only 6 known members.
We determine the luminosity
function of the group from g$^\prime$, i$^\prime$ photometry down to
i$^\prime$=22 and spectroscopy of 25 members.  Sections 2 and 3 describe
the observations, the reduction procedure and the results.  In section
4 we present a discussion.  When needed, we adopt the following values
for the cosmological parameters: H$_0$ = 70  km s$^{-1}$ Mpc$^{-1}$,
$\Omega_M$=0.3, and $\Omega_{\Lambda}$=0.7.

\section{\label{obs}Observations and reductions}

The imaging and multi-slit spectroscopic observations of the galaxy 
group RX~J1416.4+2315 were done with the GMOS instrument \citep{GMOS},
mounted on the Gemini North telescope on Feb 22th and Mar 6--7/2005
respectively.

\begin{figure}[h!]
\includegraphics[width=1.0\columnwidth]{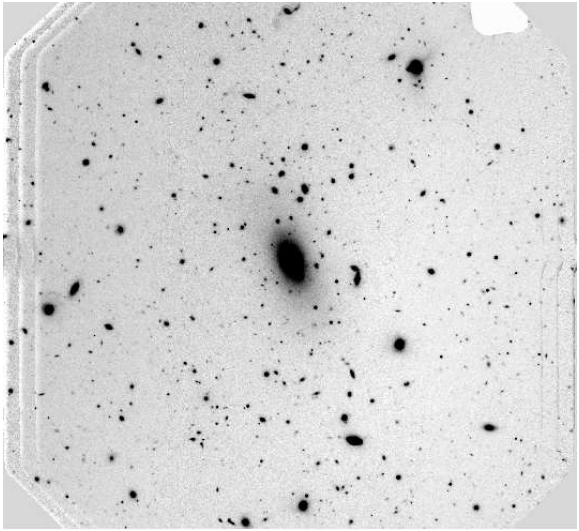}
\caption{\label{image}Optical image of RX~J1416.4+2315. 
The field of view is 5.6 arcmin 
on a side, or 814 h$_{70}^{-1}$ kpc  at the object redshift. North is
up and East is to the left.}
\end{figure}

The imaging consisted of three dithered 200s exposures in each of the two
filters from the SDSS system \citep{sloan} g$^\prime$ and i$^\prime$. The
typical FWHM for point sources was about 0.75" in all images. The
observations were performed in photometric conditions.  Fig. \ref{image}
displays the i$^\prime$ image of the system.


The calibration to the standard SDSS photometric system was made
using four stars of the \citet{landolt} field PG1323-086, which has been
calibrated to the SDSS system.
Using the dispersion between the measurements of  the 
four stars we estimated the  accuracy of the zero-point magnitude
as 0.06 in both filters.

Standard reduction using the Gemini package GMOS was used. After flat fielding
and cleaning the images for cosmic rays,  the final frames were analysed with
the program SExtractor \citep{ber96}. Positions and magnitudes (total and
aperture) were obtained for all objects. We estimate that the galaxy catalogue
is essentially complete down to 23.5 i$^\prime$ magnitude, since the number
counts turn over at i'=24 mag.

\begin{figure}[h!]
\includegraphics[width=1.0\columnwidth]{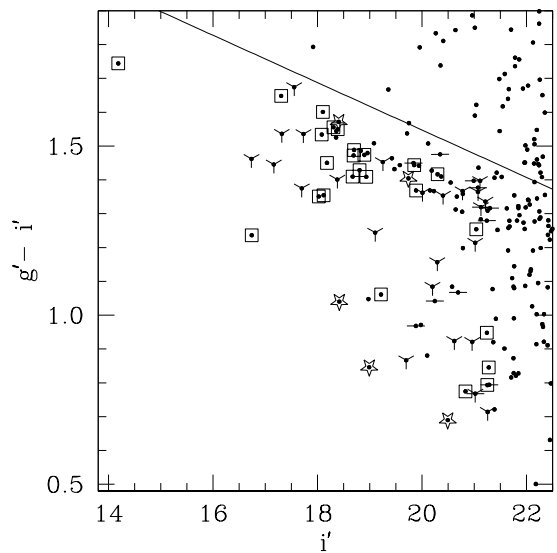}
\caption{\label{cmd}
Color-magnitude diagram of the galaxies in the RX~J1416.4+2315 field. Points
marked with squares (members), `Y' (non-members), stars (group of five galaxies
at z=0.131) and  `--' (with spectra but no redshift) represent the galaxies
observed spectroscopically. The line indicates an upper limit  for the cluster
red-sequence we adopted when selecting the spectroscopic targets. The equation
of the continuous line is  g$^\prime$ -- i$^\prime$ = --0.07 i$^\prime$ + 2.95.
Magnitudes have been measured using SExtractor's {\sc MAG\_BEST} parameter,
whereas for colors we used  3.0\arcsec~diameter apertures.}
\end{figure}

Candidates for spectroscopy were chosen based on the color-magnitude diagram
shown in Fig. \ref{cmd}: { 67 out of the 99 galaxies with M$_i$ $<$ --17.7
(m$_i<21.5$)  and bluer than the line drawn just above the red sequence
(see continuous line in Figure) were observed spectroscopically. We successfully
determined velocities of 55 of these galaxies.} Galaxies above  the
red-sequence  line are expected to be in the background, since their colors are
redder than the expected colors of elliptical galaxies at the cluster
redshift.  Note that the outermost observed galaxy which turned out to be a
member of the group/cluster has a distance of 542 h$_{70}^{-1}$ kpc from the
X-ray center of RX~J1416.4+2315.

\begin{figure}[h!]
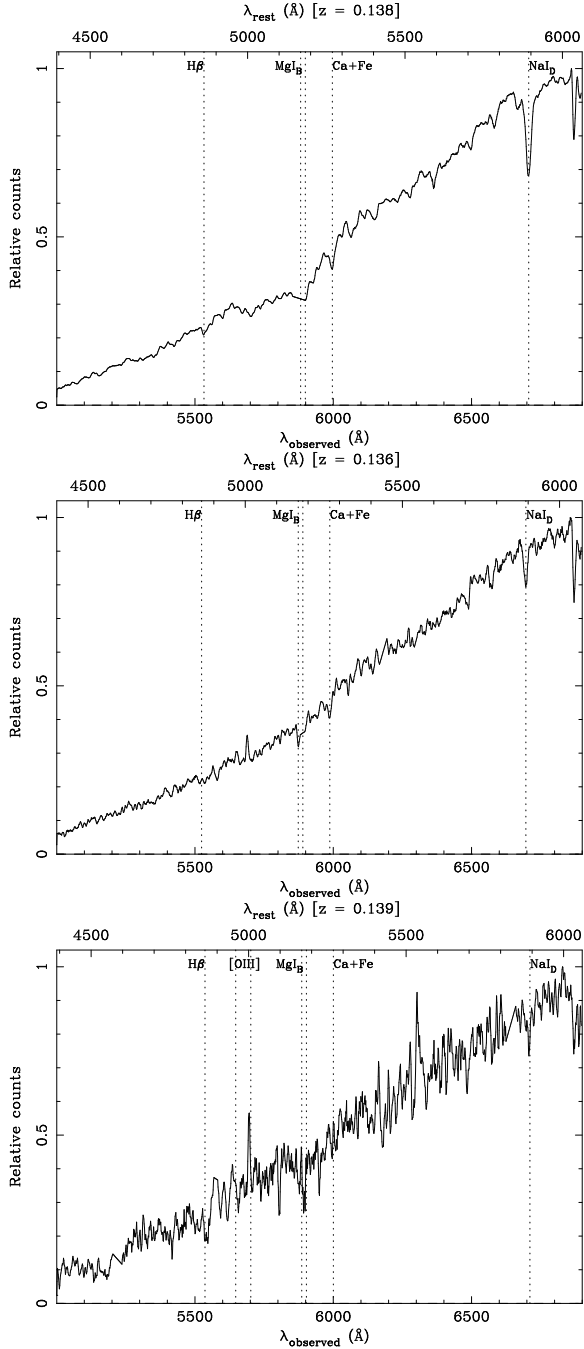

\includegraphics[width=1.0\columnwidth]{f3a.eps} \\
\includegraphics[width=1.0\columnwidth]{f3b.eps} \\
\includegraphics[width=1.0\columnwidth]{f3c.eps} 
\caption{\label{spectra}
Spectra of member galaxies. The panels show, from top to down, spectra of
galaxies G26.9+1524, G24.1+1413 and G33.6+1505 as examples of the data taken
for this study.  These spectra are the ones for the galaxies with largest,
median and lowest luminosities. respectively.  The wavelength range used for
the the cross-correlation (5000--6900\AA) is shown. In this figure,  the
spectra have been smoothed using a boxcar filter of size 13.6 \AA~(5 pixels),
for the sake of clarity.}
\end{figure}

Two multi-slit exposures of 70 minutes, each divided into three exposures, 
were obtained with GMOS through a mask with 1.0\arcsec ~ slits, using the R400
grating, for a final resolution of  6.5 \AA~(as measured from the FWHM of the
arc lines), covering approximately the range 4000 -- 8000 \AA~ (depending on
the position of each slitlet). Spectra of three of the 
members of \RXJ14 are shown in Fig. \ref{spectra}, for the galaxy with
the largest, median and lowest luminosity in the sample.


 Standard procedures were used to reduce the multi-slit spectra using tasks
within the Gemini {\sc IRAF} \footnote{IRAF is distributed by the National
Optical Astronomy Observatories, which are operated by the Association of
Universities for Research in Astronomy, Inc., under cooperative agreement with
the National Science Foundation.} package. Wavelength calibration was done
using Cu-Ar comparison-lamp exposures before and after the exposure. 

Redshifts for galaxies with absorption lines were determined using the
cross-correlation technique \citep{T&D} as implemented in the package {\sc
RVSAO} \citep{rvsao} running under {\sc IRAF}. The final heliocentric
velocities of the galaxies were obtained by cross-correlation with several
template spectra. The final errors on the velocities were determined from the
dispersion in the velocity estimates using several different galaxy and star
templates. In the case of the five emission-line redshifts, the error was
estimated from the dispersion in redshifts obtained using different emission
lines.  The resulting heliocentric velocities typically have estimated rms
errors between 19 and 78 km s$^{-1}$.  


Table \ref{tab_spectra} lists positions, total magnitudes, 
aperture (g' - i') colors, radial velocities and the \citet{T&D}
cross-correlation coefficient R for all galaxies with reliable velocity 
determination obtained in this study. 

\begin{deluxetable}{lccccccc}
\tablewidth{0pt}
\tablecaption{Properties of the galaxies in the RX~J1416.4+2315 field. \label{tab_spectra}}									   
\tablehead{\colhead{(1)} & \colhead{(2)} & \colhead{(3)} & \colhead{(4)} &
\colhead{(5)} & \colhead{(6)} & \colhead{(7)}\\
\colhead{Name} & \colhead{RA (2000)} & \colhead{DEC (2000)} &
\colhead{i$^\prime$ (AB Mag.)} &
\colhead{g$^\prime$-i$^\prime$} & \colhead{Vel. (km s$^{-1}$)} & \colhead{~R}}													   
\startdata
G28.8+1752\tablenotemark{a}   &  14 16 28.8  &   +23 17 52  &  20.40  &  1.35  &  $ ~~~759 \pm ~37$  &  3.62   \\
G28.8+1612      	      &  14 16 28.8  &   +23 16 12  &  21.21  &  1.34  &  $ ~~1375 \pm ~31$  &  3.37   \\
G22.7+1321      	      &  14 16 22.7  &   +23 13 21  &  20.62  &  0.92  &  $ ~~3967 \pm ~98$  &  3.53   \\
G18.5+1251      	      &  14 16 18.5  &   +23 12 51  &  19.25  &  1.45  &  $ ~~4799 \pm ~54$  &  3.12   \\
G17.9+1340      	      &  14 16 17.9  &   +23 13 40  &  17.73  &  1.54  &  $ ~~6554 \pm ~66$  &  3.18   \\
G24.5+1303      	      &  14 16 24.5  &   +23 13 03  &  21.07  &  1.36  &  $ ~~9966 \pm ~99$  &  3.04   \\
G24.8+1319      	      &  14 16 24.8  &   +23 13 19  &  20.96  &  0.92  &  $ ~20906 \pm ~51$  &  4.47   \\
G35.5+1353      	      &  14 16 35.5  &   +23 13 53  &  21.13  &  1.32  &  $ ~21659 \pm ~74$  &  2.86   \\
G28.9+1241      	      &  14 16 28.9  &   +23 12 41  &  17.70  &  1.38  &  $ ~25911 \pm ~86$  &  3.90   \\ 
G21.2+1722		      &  14 16 21.2  &   +23 17 22  &  16.73  &  1.46  &  $ ~30718 \pm ~88$  &  2.44   \\ 
G21.3+1724      	      &  14 16 21.3  &   +23 17 24  &  17.16  &  1.45  &  $ ~30838 \pm ~27$  &  7.79   \\
  	        	      &		     &		    &	      &	       &                               \\
G37.7+1657      	      &  14 16 37.7  &   +23 16 57  &  20.49  &  0.69  &  $ ~38851 \pm ~24$  & \nodata\tablenotemark{b}\\
G27.3+1618      	      &  14 16 27.3  &   +23 16 18  &  19.74  &  1.40  &  $ ~39112 \pm ~47$  &  3.46	\\
G36.1+1624      	      &  14 16 36.1  &   +23 16 24  &  18.42  &  1.04  &  $ ~39263 \pm ~42$  & \nodata\tablenotemark{b}\\
G34.6+1543      	      &  14 16 34.6  &   +23 15 43  &  18.99  &  0.85  &  $ ~39279 \pm ~346$ & \nodata\tablenotemark{b}\\
G27.0+1758      	      &  14 16 27.0  &   +23 17 58  &  18.41  &  1.57  &  $ ~39572 \pm ~18$  & \nodata\tablenotemark{b}\\
  	        	      &		     &		    &	      &	       &                               \\
G20.9+1605      	      &  14 16 20.9  &   +23 16 05  &  18.90  &  1.47  &  $ ~40185 \pm ~27$  &  7.78	 \\
G37.9+1712      	      &  14 16 37.9  &   +23 17 12  &  18.69  &  1.47  &  $ ~40289 \pm ~27$  &  6.29	 \\
G21.0+1711      	      &  14 16 21.0  &   +23 17 11  &  19.22  &  1.06  &  $ ~40498 \pm ~46$  &  2.53	 \\
G27.6+1607      	      &  14 16 27.6  &   +23 16 07  &  18.18  &  1.45  &  $ ~40527 \pm ~32$  &  4.28	 \\
G29.1+1544      	      &  14 16 29.1  &   +23 15 44  &  19.89  &  1.37  &  $ ~40541 \pm ~36$  &  3.62	 \\
G27.5+1548      	      &  14 16 27.5  &   +23 15 48  &  18.08  &  1.53  &  $ ~40597 \pm ~31$  &  5.80	 \\
G25.2+1541      	      &  14 16 25.2  &   +23 15 41  &  19.85  &  1.44  &  $ ~40615 \pm ~78$  &  3.25	 \\
G30.4+1518      	      &  14 16 30.4  &   +23 15 18  &  21.24  &  0.95  &  $ ~40673 \pm ~51$  & \nodata\tablenotemark{b} \\
G35.1+1442      	      &  14 16 35.1  &   +23 14 42  &  18.02  &  1.35  &  $ ~40719 \pm ~22$  &  9.36	 \\
G24.0+1332      	      &  14 16 24.0  &   +23 13 32  &  16.74  &  1.24  &  $ ~40845 \pm ~19$  & \nodata\tablenotemark{b}\\
G24.1+1413      	      &  14 16 24.1  &   +23 14 13  &  18.81  &  1.43  &  $ ~40870 \pm ~57$  & \nodata\tablenotemark{b} \\
G22.3+1729      	      &  14 16 22.3  &   +23 17 29  &  18.67  &  1.41  &  $ ~40922 \pm ~48$  &  5.17	 \\
G33.7+1733      	      &  14 16 33.7  &   +23 17 33  &  18.93  &  1.41  &  $ ~40941 \pm ~46$  &  5.75	 \\
G20.6+1517      	      &  14 16 20.6  &   +23 15 17  &  18.39  &  1.55  &  $ ~41149 \pm ~34$  &  7.98	 \\
G16.5+1601                    &  14 16 26.5  &   +23 16 01  &  18.31  &  1.56  &  $ ~41220 \pm ~43$  &  6.22	 \\
G26.9+1524      	      &  14 16 26.9  &   +23 15 24  &  14.19  &  1.74  &  $ ~41393 \pm ~43$  &  7.28	 \\
G21.7+1407      	      &  14 16 21.7  &   +23 14 07  &  21.04  &  1.25  &  $ ~41537 \pm ~55$  &  2.56	 \\	 
G22.1+1452      	      &  14 16 22.1  &   +23 14 52  &  21.25  &  0.79  &  $ ~41545 \pm ~39$  & \nodata\tablenotemark{b} \\
G33.6+1505      	      &  14 16 33.6  &   +23 15 05  &  21.28  &  0.85  &  $ ~41612 \pm ~64$  &  2.39	 \\
G28.7+1705      	      &  14 16 28.7  &   +23 17 05  &  18.12  &  1.35  &  $ ~41778 \pm ~51$  &  5.23	 \\
G24.0+1519      	      &  14 16 24.0  &   +23 15 19  &  18.10  &  1.60  &  $ ~41948 \pm ~42$  &  6.92	 \\
G17.0+1547      	      &  14 16 17.0  &   +23 15 47  &  20.83  &  0.77  &  $ ~41963 \pm ~26$  & \nodata\tablenotemark{b} \\
G23.9+1512      	      &  14 16 23.9  &   +23 15 12  &  17.30  &  1.65  &  $ ~42052 \pm ~40$  &  6.95	 \\
G31.2+1415      	      &  14 16 31.2  &   +23 14 15  &  20.30  &  1.42  &  $ ~42060 \pm ~68$  &  4.78	 \\
G34.0+1359      	      &  14 16 34.0  &   +23 13 59  &  18.70  &  1.49  &  $ ~42407 \pm ~41$  &  5.58	 \\	 
  	        	      &		     &		    &	      &	       &                               \\
G21.4+1625      	      &  14 16 21.4  &   +23 16 25  &  20.20  &  1.08  &  $ ~51515 \pm 116$  & \nodata\tablenotemark{b} \\
G37.7+1454      	      &  14 16 37.7  &   +23 14 54  &  17.32  &  1.54  &  $ ~51533 \pm 109$  & \nodata\tablenotemark{b} \\
G36.6+1506      	      &  14 16 36.6  &   +23 15 06  &  17.55  &  1.67  &  $ ~51603 \pm ~56$  &  5.00	 \\
G38.3+1308      	      &  14 16 38.3  &   +23 13 08  &  19.70  &  0.87  &  $ ~51625 \pm ~96$  & \nodata\tablenotemark{b} \\
G34.0+1737      	      &  14 16 34.0  &   +23 17 37  &  20.01  &  1.36  &  $ ~55414 \pm ~68$  &  3.74	 \\
G28.9+1448      	      &  14 16 28.9  &   +23 14 48  &  21.11  &  1.40  &  $ ~56559 \pm ~82$  &  3.17	 \\	 
G26.2+1608      	      &  14 16 26.2  &   +23 16 08  &  18.38  &  1.40  &  $ ~78740 \pm ~15$  & \nodata\tablenotemark{b} \\
G30.9+1749      	      &  14 16 30.9  &   +23 17 49  &  19.10  &  1.24  &  $ ~78798 \pm ~18$  & \nodata\tablenotemark{b} \\
G38.1+1528      	      &  14 16 38.1  &   +23 15 28  &  21.26  &  0.71  &  $ ~79061 \pm ~53$  & \nodata\tablenotemark{b} \\
G17.2+1455                    &  14 16 17.2  &   +23 14 55  &  21.02  &  0.77  &  $ ~80379 \pm ~13$  & \nodata\tablenotemark{b}\\
G36.9+1357      	      &  14 16 36.9  &   +23 13 57  &  20.78  &  1.37  &  $ 103525 \pm ~38$  & \nodata\tablenotemark{b} \\
G17.9+1624      	      &  14 16 17.9  &   +23 16 24  &  20.29  &  1.16  &  $ 156372 \pm ~53$  & \nodata\tablenotemark{b} \\
G26.4+1410      	      &  14 16 26.4  &   +23 14 10  &  21.02  &  1.21  &  $ 108704 \pm ~42$  & \nodata\tablenotemark{b} \\
\enddata
\tablenotetext{a}{The names of the galaxies are based on their 2000 celestial coordinates
(RA seconds and DEC minutes and seconds). Thus galaxy Gab.c+defg is located at
14 16 ab.c +23 de fg.}
\tablenotetext{b}{Redshift measured from emission lines.}
\end{deluxetable}	

\clearpage

\section{\label{res}Results}

\subsection {Galaxy velocity distribution}

Using the heliocentric radial velocities listed in Table \ref{tab_spectra}, we
define the putative members of RX~J1416.4+2315 as the 25 galaxies with
velocities between 40185 and 42407 km s$^{-1}$.

\begin{figure}
\includegraphics[width=1.0\columnwidth]{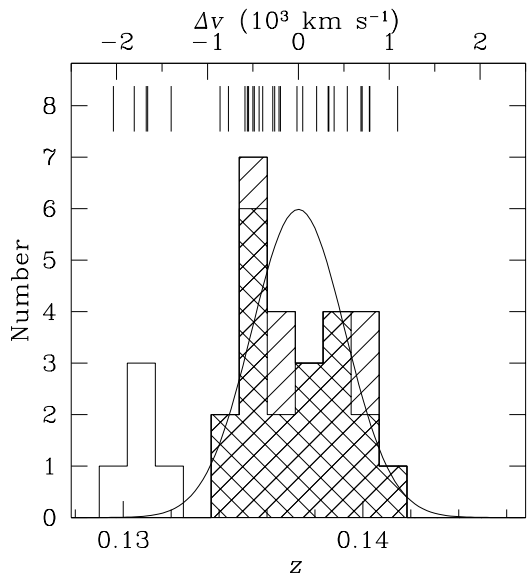}
\caption{\label{veldisp}
Velocity histogram of RX~J1416.4+2315.  It shows the distribution of the radial
velocities of 30 galaxies in the inner 542 h$_{70}^{-1}$ kpc  radius of
RX~J1416.4+2315, with redshifts within $\pm$ 2200 km/s of the systemic velocity
of the cluster. The sticks on the upper part of the plot show velocities of
individual objects.  The {\sc rostat} bi-weighted estimator gives a velocity
dispersion  $\sigma = 846$ km/s and a redshift of $\langle z \rangle = 0.1365$,
for the sample of 30 objects. The dashed and crossed areas indicate  a
sub-sample of 25 galaxies which exclude the five galaxies with lowest
velocities. The five deviant objects (indicated by the empty histogram to the
left of the figure)  are all concentrated to the northeastern side of the
cluster and probably correspond to a foreground and/or infalling group.  The
bi-weighted estimator returned $\sigma=584$ km/s and $\langle z \rangle =
0.1373$ (continuous line) for the subsample of 25 galaxies.  The crossed and
hatched areas correspond respectively to the absorption and emission-line 
members of the cluster.}
\end{figure}

Initially the sub-sample with velocities between 38851 and 42407 km s$^{-1}$
were analysed with the statistical software {\sc rostat} \citep{beers}. It
found a large gap in the velocity distribution between a foreground group of
five galaxies at a mean redshift of 0.131 and the 25 galaxies which constituted
what we identified as the cluster itself, at z = 0.137. Taking out these five
galaxies, four of which have strong emission lines, no other data points were
found outside a $\pm3\sigma$ range.  Fig. \ref{veldisp} shows the velocity
histogram for all 30 galaxies, the five objects to the left of the diagram
corresponding to the  galaxies which probably belong to the foreground or
infalling group mentioned above. These  five objects are concentrated to the
northeastern part of the cluster.


Using the robust bi-weighted estimator, {\sc rostat}, the following values for
the systemic redshift and velocity dispersion were found: $\langle z \rangle =
0.1373 \pm0.0009$ and $\sigma = 584\pm121$ km s$^{-1}$, for the 25 member
galaxies, excluding the quintet, which has a mean velocity $\sim$ 2000 km
s$^{-1}$ below the mean velocity of the cluster itself.

The inclusion of the five objects increases the velocity dispersion to $\sigma =
846\pm321$ km s$^{-1}$, for a mean redshift $\langle z \rangle = 0.136\pm0.001$.
We consider the determination of redshift and velocity dispersion for the 25
members  the most representative values for RX~J1416.4+2315, given the large
velocity  difference between the foreground group and the cluster. Note that for
the determination of the luminosity function only the 25 galaxies were used.

We also calculated the velocity dispersion of \RXJ excluding five galaxies with
strong emission lines from the sample of 25 members, since those galaxies often
are not in dynamical equilibrium with the whole structure
\citep[e.g.][]{sodre89,biviano97}. By doing this  exercise we found essentially
the same results for the redshift ($\langle z \rangle=0.137$) and the velocity
dispersion  ($\sigma=605$ km s$^{-1}$).

Using the  $R_{200} = {\sqrt{3} \sigma_{cl} \over 10 H(z)}$ \citep{CNOC}, as an
estimator of the virial radius, we find a value of 1.2 Mpc for the  sample with
25 galaxies. The spectroscopy presented here is then confined to about one half
of the virial radius.

\subsubsection{Dynamical mass \label{dymass}}

We determine the dynamical mass of the cluster by using four different mass
estimators, as suggested by \citet{heisler85}: virial, projected, average and
median mass estimators (see Table \ref{mass}).  The final, adopted mass for the
cluster, M = 1.9 $\times$ $10^{14}$ \Msol, was obtained from the average value
of the results of the four different estimators. { This result agrees with
the mass estimated from X-ray observations done recently by \citet{habib06},
within a confidence level of 68\%.}

We should keep in mind the caution notes given by several authors
\citep[e.g.][]{girardi} on the reliability of mass determinations when the
galaxies are distributed over a small (central) part of the cluster and the
number of redshifts is limited. 

\begin{deluxetable}{lcccc}
\tablewidth{0pt}
\tablecaption{Mass Estimates \label{mass}}									   
\tablehead{\colhead{(1)} & \colhead{(2)} & \colhead{(3)} & \colhead{(4)} &
\colhead{(5)}\\
\colhead{} & \multicolumn{2}{c}{25 galaxies (z=0.137)} &
\multicolumn{2}{c}{All 30 galaxies (including group)} \\											   
\colhead{Estimator} & \colhead{Mass ($10^{14}$ \Msol)} &
\colhead{M/L$_B$ (h$_{70} $\Msol/L$_B$\sol)} & \colhead{Mass ($10^{14}$ \Msol)} &
\colhead{M/L$_B$ (h$_{70}$ \Msol/L$_B$\sol)}\\}
\startdata
Virial     & 1.9 & 450 & 4.8 & ~996 \\
Projected  & 2.4 & 571 & 6.8 & 1404 \\
Average    & 1.8 & 427 & 4.2 & ~873 \\
Median     & 1.4 & 331 & 3.7 & ~768 \\
           &     &     &     &      \\
Mean value & 1.9 & 445 & 4.9 & 1010 \\	      
\enddata
\end{deluxetable}

\subsubsection{Mass-to-light ratio} 
  
We estimated the luminosity of the central regions of the cluster by adding up
the luminosities of the spectroscopically confirmed members, taking into account
the completeness correction derived from  the spectroscopic sampling (see Fig.
\ref{flum}).   The completeness correction was defined in \citet{paper1}. For
this analysis, we have neglected all the galaxies redder than the limit of the
red-cluster sequence, drawn in Fig. \ref{cmd}.

In order to compare with previous results, we have transformed our SLOAN
g$^\prime$ magnitude in standard Johnson--Morgan B magnitudes by adopting  a
color  $g^\prime - B = 0.634$ from \citet{fukugita95}. This is the typical color
for an S0 galaxy, which we adopt as being representative of the morphological
mix found in \RXJ, interpolated to the redshift of the cluster.  The total
magnitudes have been corrected for Galactic--extinction \citep{schlegel} as well
as for the $k$-correction    \citep{fukugita95}, under the assumption that all
galaxies have early morphological types, which is valid for  most of the sample.

The total luminosity calculated within 542 kpc h$_{70}^{-1}$ is  $4.42\times
10^{11}$ h$_{70}^{-2}$ L$_{\odot B}$ for 25 galaxies (and $4.83\times 
10^{11}$ h$_{70}^{-2}$ L$_{\odot B}$ if we
include the foreground quintet; stars in Fig. \ref{cmd}). This  leads to a
mass-to-light ratio in units of h$_{70}$ M\sol/L$_B$\sol ~ of 445. Results for the
several mass estimates  including or not the foreground group  are presented in
Table \ref{mass}.

{  The velocity dispersion, dynamical mass, mass-to-light ratio and the
richness of the structure  determined from
our observations, are typical of clusters. Taking also into account the X-ray
measurements of \citet{habib06},  we can safely characterise RX~J1416.4+2315 as a
cluster. }

\subsection{The Luminosity Function}

\begin{figure*}[htq]
\includegraphics[width=1.0\columnwidth]{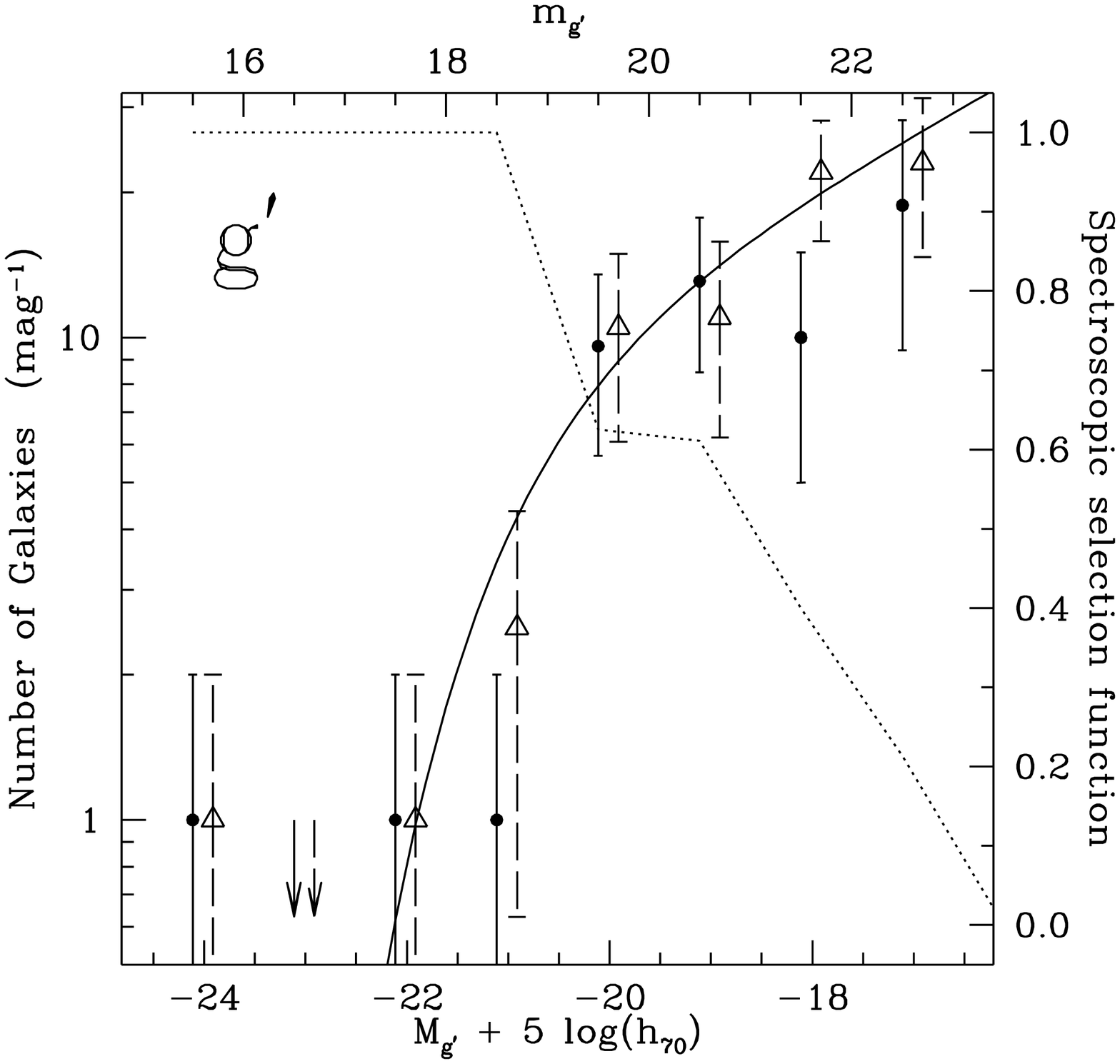}
\includegraphics[width=1.0\columnwidth]{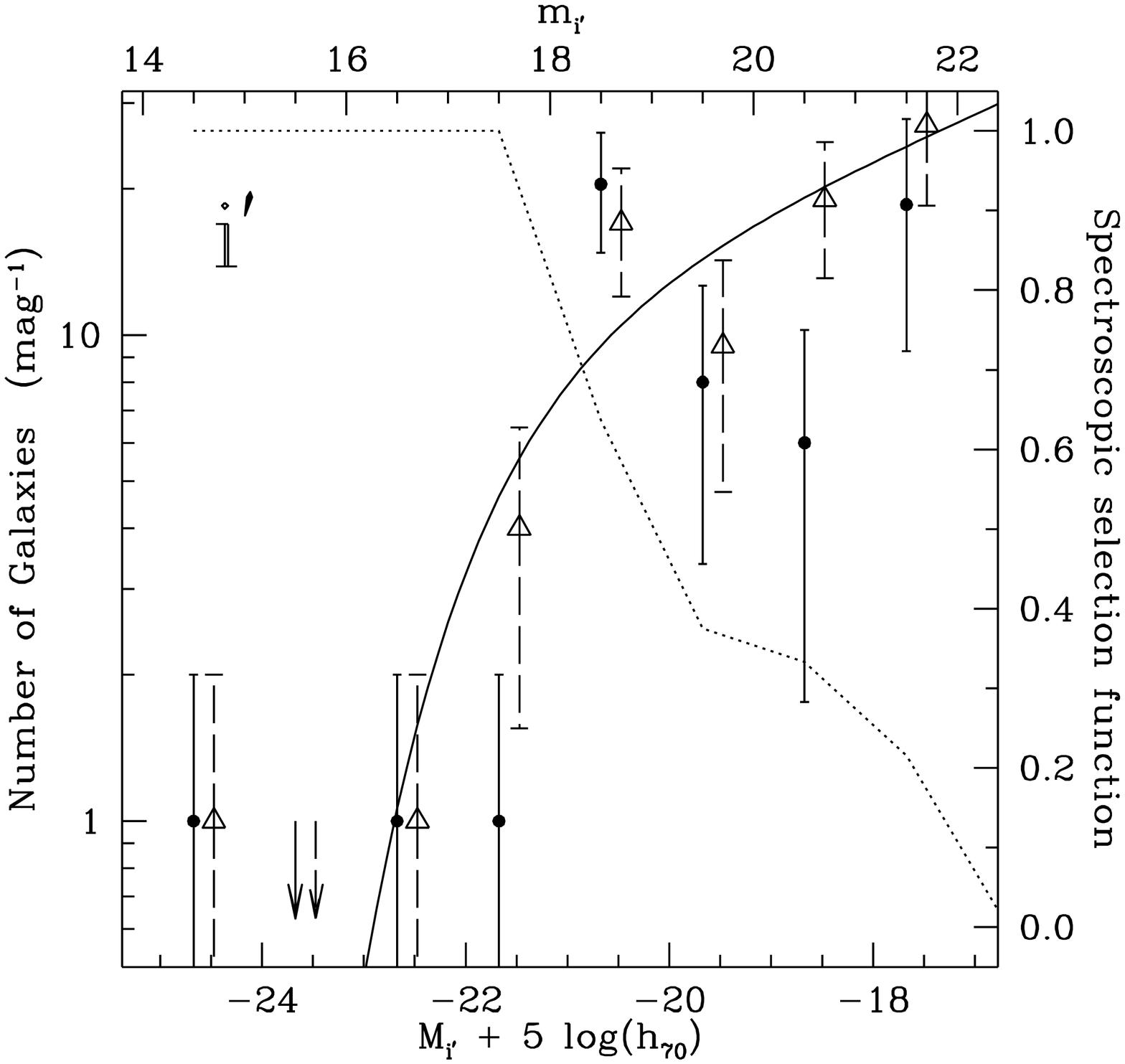}
\caption{\label{flum}
Luminosity Function of RX~J1416.4+2315.  The panels show, from left to right,
the luminosity functions in the g$^\prime$ and i$^\prime$ bands, respectively.
The solid circles show the completeness-corrected number of spectroscopically
confirmed members of  RX~J1552 per 1.0 magnitude bin in the GMOS field. The
error bars are 1$\sigma$ Poissonian errors. The arrows show bins with number of
galaxies less or equal to zero. The dotted line is the selection function of 
the
spectroscopic sample. The open triangles show the photometrically-determined 
luminosity function estimated through number counts and statistical subtraction
of the background.  The points have been shifted by 0.2 mag, for the sake of
clarity. The continuous lines show the best fitted Schechter functions to the
photometric luminosity  functions (see values in Table \ref{tab_FdL}).}
\end{figure*}

We show in Fig. \ref{flum} the luminosity function  (LF) of RX~J1416.4+2315 
(solid circles)  obtained with the  25 galaxies with spectroscopically confirmed
membership. The absolute magnitudes were calculated after correcting the
observed magnitudes for Galactic extinction and applying $k$-corrections. The
selection function, also shown in Fig. \ref{flum}, was calculated considering
only galaxies bluer than the upper limit of the adopted red cluster sequence. 
We have also estimated photometrically the LF of  RX~J1416.4+2315 
down to 22 and 23 mag for
the i$^\prime$ and g$^\prime$ bands respectively, comfortably within the
completeness limit, by adopting the procedure described in \citet{paper1}, which
makes use of  empty control fields imaged with the same instrument and filters
(Boris et al. 2006, in preparation) but taken with longer exposure times. The
photometric luminosity functions are shown in Fig. \ref{flum} as open
triangles. 

\begin{deluxetable}{cccc}
\tablewidth{0pt}
\tablecaption{Photometric Luminosity Function \label{tab_FdL}}									   
\tablehead{\colhead{(1)} & \colhead{(2)} & \colhead{(3)} & \colhead{(4)} \\
 \colhead{Band} & \colhead{M$^*+5\log(h_{70})$} &
\colhead{$\alpha$} & \colhead{$\langle M\rangle+5\log(h_{70})$}  }
\startdata
g$^\prime$ & $-21.16\pm0.76$  & $-1.23\pm0.24$ & 17.0 -- 23.0\\
i$^\prime$ & $-21.86\pm1.01$  & $-1.29\pm0.27$ & 16.0 -- 22.0\\
\enddata
\end{deluxetable}


\clearpage

It can be seen from Fig. \ref{flum} that both ways of estimating the LF  
lead to the same qualitative results, but clearly the spectroscopically
determined LF is noisy at the faint end (mainly in the
i$^\prime$ band) and it does not fit a Schechter function well. On the other
hand, the photometric LFs are  better determined and they do fit Schechter
functions.  In Table \ref{tab_FdL} we show the best fitted parameters of a
Schechter function fitted to the photometric LFs. 

The results in both bands are similar and they show a large gap between the two
brightest galaxies of the system, which characterises a fossil system. At the
bright end, there are very few galaxies with magnitudes close to M$^*$. At the
faint end, there is a significant increase of the number of dwarf galaxies, with
the best fit being for a Schechter parameter $\alpha$ between -1.2 and -1.3,
which is typical of clusters \citep{propris,popessoII}.

{ Regarding the faint end slope, the results presented here are very
consistent with the spectroscopic selected LF of \RXJ~ 
shown in \citet[][$\alpha = -1.23$]{habib06}, but not with their photometric selected LF
($\alpha = -0.61$). }

\subsection{Surface photometry of the brightest cluster galaxy}

\begin{figure}[!ht]
\includegraphics[width=\columnwidth]{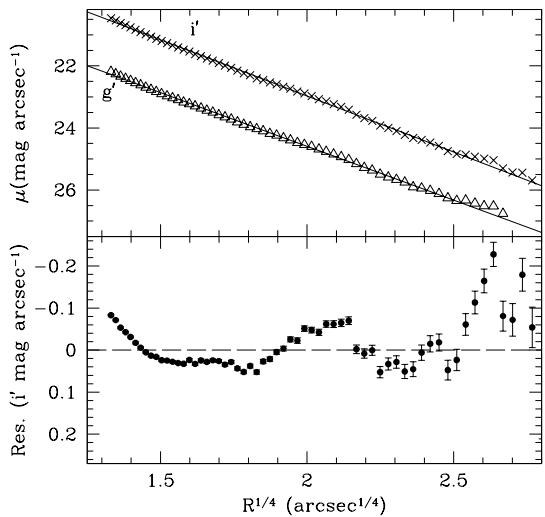}
\caption{\label{phot} {\it Upper panel} -- Photometric profile of the central
galaxy. We show the surface brightness profile between a value of the semi-major
axis of  2.5\arcsec ~
and that where the counts reach 1$\sigma$ of the background
level (26.8 and 25.8 mag arcsec$^{-2}$ for g$^\prime$, and i$^\prime$ bands
respectively) as a function of semi-major axis to the power 1/4. The solid line
is the best fit de Vaucoleurs profile between 3.0 and 28.2\arcsec ~
($\mu_{i^{\prime}}=24.0$ mag arcsec$^{-2}$). {\it Lower panel} -- Residual
between the actual r$^\prime$ band profile and the de Vaucoleurs profile fit.}
\end{figure}

In the upper panel of Fig. \ref{phot}, the azimuthally averaged photometric
profile of the central galaxy of RX~J1416.4+2315 is shown.   The surface
photometry was performed using the task {\sc ELLIPSE} in {\sc STSDAS/IRAF},
which fits ellipses to extended object isophotes. We allowed the ellipticity and
position angle of the successive ellipses to change but the center remained
fixed.  The ellipse fitting was performed only in the higher signal to noise
i$^\prime$ image. For the g$^\prime$ band, the software measured the isophotal 
levels using the parameters estimated in the i$^\prime$ image. There were a few
small objects within the outer isophotes of the  central galaxy which were
masked during the profile fitting procedure.


We have fitted an r$^{1/4}$-profile to the bright end of the galaxy surface
brightness profile,  from well outside the seeing disk, at 3.0\arcsec, to a
semi-major axis of 22.8\arcsec, which corresponds to the ellipse  that contains
half of the total galaxy luminosity and has a surface brightness
$\mu_{i^{\prime}}=24.1$ mag arcsec$^{-2}$. In the lower panel of Fig.
\ref{phot}, the residuals (data minus r$^{1/4}$-profile model) for the
i$^\prime$-band data are shown. We find no significant  light excess over the
de-Vaucoleurs profile, in either band. This result agrees with previous analysis
done by \citet{jones03}. It is in contrast with the significant light excess
found for the  brightest galaxy of \RXJold.

\section{Discussion}

It is enlightening to compare the results presented here with the ones found for
\RXJold~, the only other fossil system which has been studied spectroscopically
so far.  Both are at similar redshifts and we have studied them using data from
the same telescope and setup and using the same analysis methodology.  Both
systems were originally thought to be fossil groups (before our study) but
turned out to be  fossil clusters instead, as shown in Section \ref{res} and
summarised below.

Despite \RXJ having a bolometric X-ray emission of about three times that of
\RXJold, both systems have  comparable velocity dispersion ($\sim 600$ km
s$^{-1}$), dynamical mass ($\sim 2\times 10^{14}$ h$^{-1}_{70}$ 
M$_\odot$ within about one
half of the virial radius) and  mass-to-light ratio (450--500 
h$_{70}$ M\sol/L$_B$\sol).
Taking these values into account, plus the richness of the structures, both
systems can be classified as fossil clusters, which, following
\citet{chrismiller}, put them among very few other objects known
in the Universe.

The luminosity functions of the two systems are not similar, however. While
\RXJold\\ presented a decreasing number of faint galaxies ($-19 < M_{g^\prime} <
-17 $), being well fitted by a Schechter function with  $\alpha$ = --0.4 to
--0.8, 
\RXJ presents an increasing one with $\alpha$ = --1.2 to --1.3. Regarding
the inclination of the luminosity function in this magnitude interval, \RXJ is
very consistent with that observed for the average cluster population
\citep{propris,popessoII}.  It is interesting to note that for \RXJold we found
an extended light envelope around the central galaxy of the system, measured as
an excess above a de Vaucoleurs profile, which indicated that the central object
was a cD galaxy. Mendes de Oliveira et al. (2006) argued that the envelope
around the cD may be related to the lack of faint galaxies. In that case and
also in \citet{Omar} it was argued that tidal disruption of faint galaxies can
form such a halo { \citep[see also][]{Cypriano06}}.  For \RXJ we did not find
a lack of faint galaxies neither a light envelope associated with the central
galaxy, which is consistent with the scenario advocated in these previous
studies. { Note that one might attribute the differences between the
two fossil clusters to differences in their masses (e.g., Lin \& Mohr 2004).
However, the small mass diference between these systems, $\sim 35$\%, 
does not seem to favor such an interpretation.}

      At the bright end of the luminosity function, \RXJold contains fewer
bright galaxies with $M<-21.5~i^\prime$ mag than \RXJ. While \RXJold~ has 7
spectroscopically confirmed members RX~1416J has only 3, in a magnitude range
where the spectroscopic completeness are close  to 100\% in both cases.  We
could speculate that if both clusters started their evolution with similar
numbers of  bright galaxies, in the case of \RXJ more members have merged to
form the central object. If this scenario is correct we would say that \RXJ,
compared to \RXJold, is in a more advanced state of fossilisation.


\begin{acknowledgements}

We are grateful to Pamela Piovezan for helping in the early phases of the
reduction of the spectroscopic data, Natalia Boris for providing the control
fields, Florence Durret for very useful suggestions, and the Gemini staff for
obtaining the observations.  { We thank Habib Khosroshahi for sharing 
his results on \RXJ prior to publication.} The authors would like also to
acknowledge support from the Brazilian agencies FAPESP (projeto tem\'atico
01/07342-7), CNPq and CAPES. We made use of the NASA/IPAC Extragalactic Database
(NED), which is operated by the Jet Propulsion Laboratory, California Institute
of Technology, under contract with NASA.

\end{acknowledgements}


\bibliographystyle{aa}
\bibliography{abb,all,new}


\end{document}